\documentclass[prb,floatfix,showkeys,footinbib,twocolumn]{revtex4-1}
\usepackage[a4paper,top=1.7cm,bottom=1.7cm,left=1.7cm,right=1.7cm]{geometry}
\usepackage{amssymb}
\usepackage{amsmath}
\usepackage{amsfonts}
\usepackage{color}
\usepackage[normalem]{ulem}
\usepackage[pdftex,colorlinks=true,citecolor=red,linkcolor=red]{hyperref}
\usepackage[force]{feynmp-auto}
\usepackage[utf8]{inputenc}
\usepackage{graphicx}
\setcitestyle{super}
\newcommand{\grad}{\mbox{\boldmath$\nabla$}}
\newcommand{\curl}{\mbox{\boldmath$\nabla$}\times}
\newcommand{\eps}{\varepsilon}
\newcommand{\onehalf}{\frac{\mbox{\small 1}}{\mbox{\small 2}}}
\newcommand{\tinyonehalf}{\frac{\mbox{\tiny 1}}{\mbox{\tiny 2}}}
\newcommand{\tinyonethird}{\frac{\mbox{\tiny 1}}{\mbox{\tiny 3}}}
\newcommand{\be}{\begin{equation}}
\newcommand{\ee}{\end{equation}}
\newcommand{\ber}{\begin{eqnarray}}
\newcommand{\eer}{\end{eqnarray}}
\newcommand{\swave}{$^{1}$S$_{0}$\,}
\newcommand{\pwave}{$^{3}$P$_{2}$\,}
\newcommand{\kb}{k_{\text{B}}}         
\newcommand{\He}{$^3$He}
\newcommand{\Hefour}{$^4$He}
\newcommand{\vOmega}{\pmb\Omega}
\def\dthree#1{\frac{d^3{#1}}{(2\pi\hbar)^3}}
\def\der#1#2{\mbox{$\displaystyle\frac{d #1}{d #2}$}}
\def\nicefrac#1#2{\genfrac{}{}{}{1}{#1}{#2}}
\def\vl{{\bf l}}
\def\vr{{\bf r}}
\def\vR{{\bf R}}
\def\vQ{{\bf Q}}
\def\vp{{\bf p}}
\def\vu{{\bf u}}
\def\vv{{\bf v}}
\def\vw{{\bf w}}

\def\vJ{{\bf J}}
\def\vB{{\bf B}}
\def\cY{{\mathcal Y}}
\def\cA{{\mathcal A}}
\def\cC{{\mathcal C}}
\def\ns{\negthickspace}

\def\urca{\text{Urca}}
\def\murca{\text{mUrca}}
\begin{document}
\title{Generation of Quantum Turbulence by Neutrino Cooling in Neutron Stars}
\author{J. A. Sauls}
\email{sauls@lsu.edu}
\affiliation{Hearne Institute of Theoretical Physics, 
             Louisiana State University, Baton Rouge, LA 70803}
\date{\today}
\begin{abstract}
The interior crust and much of the liquid core of neutron stars is believed to be a quantum liquid mixture of neutron and proton superfluids and a relativistic electron liquid. Quantized vortices in the neutron superfluid and quantized flux lines in the proton superconductor are topological defects of these hadronic condensates.
I consider the formation of the superfluid state in young neutron stars under nonequilibrium conditions imposed by the neutrino cooling rate. The nonequilibrium phase transition implies that the onset of superfluidity is accompanied by the generation of quantized vortices based on the mechanism envisioned by Kibble in the context cosmic string formation in evolutionary models of an expanding universe, and further developed by Zurek for nonequilibrium phase transitions in quantum liquids such as \Hefour.
I discuss the Kibble-Zurek mechanism (KZM) and scaling relations for topological defect formation starting from the Cooper pair fluctuation propagator for temperatures approaching $T_c$. I then calculate the predicted vortex densities based on Urca and modified Urca cooling mechanisms in the cores of neutron stars for several models of the superfluid gap and transition temperature of the interior neutron superfluid. 
In all cases studied the KZM leads to a large density of topological defects in the condensate phase, which in 3D form a random network of vortex lines and loops, i.e. the generation of quantum turbulence.
\end{abstract}
\maketitle
\vspace*{-15mm}
\section{Introduction}
\vspace*{-5mm}

Considerations of the effects of superfluidity on the thermodynamic and transport properties of nuclear matter in neutron stars often start from a superfluid state described by quasiequilibrium conditions, i.e. a condensate of Cooper pairs forms at the transition temperature, $T_c$, which opens a gap in the neutron excitation spectrum, and a condensate that is locally homogeneous except for a dilute array of nearly rectilinear vortices, separated by distances of order microns,\footnote{This scale is set by the Feynman-Onsager relation for the mean areal density of vortices as discussed in App.~\ref{sec-vortex_density-rotation}.} i.e. macroscopic distances compared to the nuclear scale of $a=1/\sqrt[3]{n}\approx 1\,\mbox{fm}$.

Rapid cooling of the nuclear matter by neutrino emission leads to a very different superfluid phase that persists and evolves over relatively long timescales, a phase, or phases, in which the condensate is highly inhomogeneous and perforated by a relatively dense tangle of topological defects in the form of quantized vortex lines and vortex rings. 
Such a state exhibits quantum turbulence in the form of space-time fluctuations of the superfluid velocity and motion of the line singularities. 

The mechanism for generating topological defects by cooling through a second-order phase transition at a finite rate into a broken symmetry state was proposed by Kibble as mechanism for generating cosmic strings as topological defects and sources for inhomogeneity in the early evolution of the universe.~\cite{kib76,kib80,vilenkin94,hin95} 
This idea was pursued by Zurek who made the connection between defect formation in symmetry breaking phase transitions in cosmological models of the early universe with nonequilibrium phase transitions in condensed matter systems such as liquid Helium.~\cite{zur85,zur96}
In the context of neutron stars the authors of Ref.~\onlinecite{bag24} posit the Kibble-Zurek mechanism (KZM) in superfluid pulsars and argued that it leaves a temporal signature on the pulsar timing curve.

In the following I consider the KZM for the nonequilibrium conditions imposed by neutrino cooling on the 
superfluid transition of neutron matter in the cores of young neutron stars, develop the KZ scaling 
relations from the Cooper pair fluctuation propagator for temperatures near $T_c$ subject to neutrino 
cooling, calculate vortex line densities in the inner and outer core of neutron stars at the 
time of formation, Kibble-Zurek freeze-out. I also discuss the generation and potential impact of 
noncurrent carrying topological line and point defects generated by the KZM in the \pwave\ core 
superfluid as well as the inhomogeneity of the transition temperature and 
neutrino rate on the generation and coarsening of quantum turbulence.

\vspace*{-7mm}
\subsection{Neutrino cooling processes}
\vspace*{-5mm}

Neutron stars are born ``hot'', with interior temperatures of order $10^{13}\,\mbox{K}$. They cool rapidly by neutrino emission processes below the neutron degeneracy temperature of order $T_{f_n}\approx 10^{12}\,\mbox{K}$.
Degenerate nuclear matter at densities of order nuclear matter density and above is predicted to be particularly efficient in generating neutrinos via beta decay in neutron rich matter, the ``direct Urca process'', (e.g. $n\rightarrow p+e^-+\bar\nu_e$),~\cite{bog81,lat91} beta decay of nucleon quasiparticles in the presence of a pion condensate,~\cite{max77,yak99} or beta decay of hyperons in the core.~\cite{pra92,pet92}
The neutrino luminosity generated by the direct Urca process, if kinematically allowed will dominate; it varies smoothly as function of the neutron density on the scale of nuclear matter density, $n_0=0.15\,\mbox{fm}^{-3}$, and scales with temperature as,
\be
L^{\urca}_{\nu} = R^{\urca}\,(\kb T)^6
\,,
\ee
where $R^{\urca}$ depends on the transition matrix element of the weak-interaction for beta decay of neutrons.~\cite{lat91,pet92} See App.~\ref{app-neutrino_cooling} for relevant details.
The proton concentration in the core must exceed a critical threshold, $x=n_p/n>x_c\approx 0.11-0.15$, for neutrino emission via the Urca reaction. The authors of Ref.~\onlinecite{lat91} noted that $x$ depends on the nuclear symmetry energy, and that for a number of viable equations of state for $1.4 M_{\odot}$ and maximum mass neutron stars the proton concentration exceeds the threshold for neutrino emission by the direct Urca process for $n\gtrsim 2n_0$.~\cite{pra88}

The cooling rate for matter in the neutron star (NS) interior is then $C_v\,dT/dt = -L_{\nu}^{\urca}$, where the specific heat of degenerate nuclear matter above the superfluid transition is dominated by neutron excitations, $C_v=\gamma_n\,T$, with Sommerfeld coefficient $\gamma_n=k_{f_n}\,M_n^*\,\kb^2/3\hbar^2$, where $M_n^*$ is the neutron quasiparticle effective mass. 
For the direct Urca process the rate scales as
%
\be
-\frac{1}{T}\der{T}{t}
=K^{\urca}\,
\left(\frac{n_0}{n}\right)^{\nicefrac{1}{3}}\,
\left(\frac{T}{10^9\,\mbox{K}}\right)^4
\equiv 
\frac{1}{\tau^{\urca}} 
\,,
\label{eq-tau_Urca}
\ee
%
with $K^{\urca}$ defined in Appendix Eq.~\eqref{eq-K_Urca}. For $n = 2n_0$, $M_p^* = 0.7\,M_p$~\cite{bal07} and $\mu_e=100\,\mbox{MeV}$ ~\cite{max77} $K^{\urca} = 5.53\times 10^{-3}\,\mbox{sec}^{-1}$, and $\tau^{\urca}=3.8\,\mbox{min}$.
Thus, if the Urca process is operable then much of the interior of the neutron star is expected to cool to $T=10^9\,\mbox{K}$ in a matter of minutes provided superfluidity does not intervene.~\cite{pet92}

If the direct Urca process is kinematically prohibited, Bose-Einstein condensation of $\pi^-$ mesons, predicted to as a novel state of nuclear matter high densities,~\cite{mig73,bar73,bay75,wei75} opens a channel for beta decay of nucleon quasiparticles, $u\rightarrow u'+e+\bar\nu_e$, where $u,u'$ are linear superpositions of neutron and proton particle and hole excitations.~\cite{bah65b,max77} 
A key feature of $\pi^-$ condensation is that it does not open a gap the fermionic excitation spectra. As a result Fermi statistics for beta decay of $u$ quasiparticles predicts the same $T^4$ dependence of the cooling rate as that for the direct Urca process. 
However, the $\pi^-$ condensate-assisted beta decay cooling rate is reduced relative to the direct Urca process with $K^{\urca}\rightarrow K^{\pi}\approx\theta_{\pi}^2\,K^{\urca}$, where $0<\theta_{\pi}\le 1$ is the $\pi^-$ condensate amplitude estimated to be of order $\theta_{\pi}^2=0.1$.~\cite{wei75} 
Similarly, if instead $\Lambda$ or $\Sigma^-$ hyperons are created in the core they too can undergo beta decay, e.g. $\Lambda\rightarrow p+e^- +\bar\nu$, and with a nearly vanishing threshold concentration. The cooling rate for hyperon beta decay is suppressed by a factor $\theta_c^2\approx 0.05$ relative to the direct Urca rate, $K^{\urca} \rightarrow K^{\Lambda} \approx \theta_c^2\,K_{\urca}$, where $\theta_c=0.2272$ is the Cabbibo angle.~\cite{pra92,pet92}  
Still these are fast cooling rates compared to modified Urca processes.~\cite{yak99,pet92,yak04} 

\begin{figure}[t]
\includegraphics[width=0.45\textwidth]{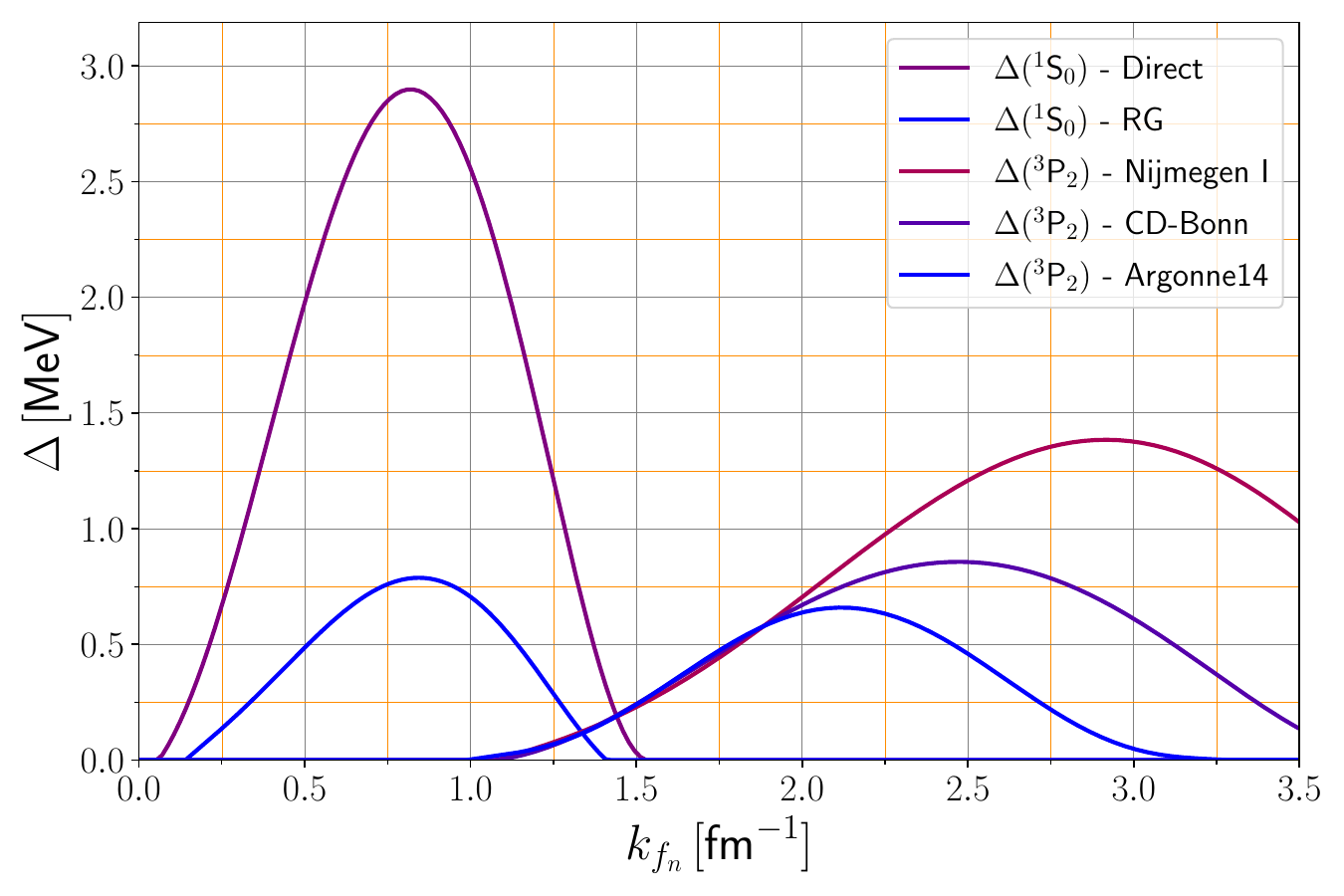}
\caption{
Neutron superfluid gaps vs neutron Fermi wavenumber. Shown are calculated gaps for \swave\ pairing at densities above neutron drip up to $k_{f_n}\approx 1.5\,\mbox{fm}^{-1}$ based on direct nuclear potentials reported in Ref.~\onlinecite{heb07}. The reduced \swave\ pairing gap is based on the one-loop renormalization group equations for interacting Fermi liquids from Ref.~\onlinecite{sch03}.
At higher densities \pwave\ superfluidity is predicted with somewhat reduced pairing gaps in the inner core.~\cite{bal98} 
N.B. The curves were generated from digitized data extracted from Fig.~2 of Ref.~\onlinecite{heb07}, Fig.~8 of Ref.~\onlinecite{sch03} and Fig.~4 of Ref.~\onlinecite{bal98}.  
}
\label{fig-Delta_vs_kfn}
\end{figure}

In the outer core and the inner crust with a lower density neutron liquid, $n < 2\,n_0$, these direct Urca processes are absent. Neutrino emission proceeds by a modified Urca process: beta decay assisted by the addition of a spectator neutron, e.g. $n+n\rightarrow n+p+e^-+\bar\nu$. The spectator neutron in the initial and final states of this reaction leads to a reduction in phase space for the neutrino luminosity by a factor of order $(\kb T / E_{f_n})^2$, and a corresponding reduction in the cooling rate compared to the direct Urca rate (if allowed)
%
\be
-\frac{1}{T}\der{T}{t}
=K^{\murca}\,
\left(\frac{n_0}{n}\right)^{\nicefrac{1}{3}}\,
\left(\frac{T}{10^9\,\mbox{K}}\right)^6
\equiv 
\frac{1}{\tau^{\murca}} 
\,,
\label{eq-tau_mUrca}
\ee
with $K^{\murca}=\left(E_9/E_{f_n}\right)^{2}\,K^{\urca}$, where
$E_9\equiv\kb\,10^9\mbox{K}\approx 0.0862\,\mbox{MeV}$ and 
$E_{f_n}=(\hbar k_{f_n})^2/2M_n^*$ is the Fermi energy of the neutron liquid.
For $k_{f_n}=1.0\,\mbox{fm}^{-1}$, $M_n^*/M_n = 1$, $M_p^*/M_p = 1$, $\mu_e=50\,MeV$ we obtain $K^{\murca}=1.71\times 10^{-8}\,\mbox{sec}^{-1}$ and $\tau^{\murca} = 412\,\mbox{days}$.


Neutrino emission by Urca and modified Urca processes are suppressed if a gap in the neutron excitation spectrum opens as a result of BCS pairing, and that is widely predicted to happen, both in the inner crust and throughout the core of neutron stars. The specific heat is also suppressed at temperatures $T \ll T_c$. Thus, the opening of a gap in the neutron and/or proton excitation spectrum leads to significant deviations from the Urca cooling rate for non-superfluid nuclear matter.
For discussion of the impact of superfludity \emph{on neutrino emission and cooling rates} see Refs.~\onlinecite{yak99,pag09}

\vspace*{-7mm}
\subsection{Pairing gaps \& transition temperatures}
\vspace*{-5mm}

At relatively low densities above neutron drip, $0.1\,\mbox{fm}^{-1}<k_{f_n}\lesssim 1.5\,\mbox{fm}^{-1}$, the neutron liquid in neutron stars is predicted to form a BCS condensate of spin-singlet, s-wave ($^1S_0$) Cooper pairs, with pairing gaps (transition temperatures) in the range $\Delta_n=0.5-3.0\,\mbox{MeV}$ ($T_{c_n}=\Delta_n/1.78$).~\cite{mig60,gin65}
At higher densities, $k_{f_n}\gtrsim 1.5\,\mbox{fm}^{-1}$, superfluidity is predicted to occur in a spin-triplet, p-wave channel with $J=2$ ($^3P_2$),~\cite{hof70,tam70} albeit with somewhat smaller pairing gaps of order $\Delta_n=0.2-1.5\,\mbox{MeV}$ ($T_{c_n}\approx\Delta_n/1.2$).
Figure~\ref{fig-Delta_vs_kfn} shows theoretical calculations of pairing gaps based on more recent nuclear potentials, all of which predict nearly identical pairing gaps at densities below $k_{f_n}=2.0\,\mbox{fm}^{-1}$, i.e. approximately twice nuclear saturation density, but lead to significantly different pairing gaps at higher densities.~\cite{heb07,bal98}
The pairing gaps for $k_{f_n}\gtrsim 2.0\,\mbox{fm}^{-1}$ are based on nuclear interactions that are extended to densities where nuclear models are poorly understood.
The strong reduction in the \swave\ pairing gap shown for $k_{f_n}\lesssim 1.5\,\mbox{fm}^{-1}$ is based on the one-loop renormalization group calculation for pairing in interacting Fermi liquids by Ref.~\onlinecite{sch03} with includes many-body corrections to the direct pairing interaction.

For $\Delta_n$ spanning $0.1-1.0\,\mbox{MeV}$, the onset of superfluidity in the core varies locally by an order of magnitude: $T_{c_n}\approx 10^{9} - 10^{10}\,\mbox{K}$.
Thus, superfluidity in neutron star cores is expected to onset early in the thermal evolution of a neutron star born from a supernova event during a period of rapid neutrino cooling.

\vspace*{-7mm}
\subsection{Nonequilibrium Phase Transition}
\vspace*{-5mm}

Here I address the impact of cooling by neutrino emission on the nature of the superfluid state that develops under such nonequilibrium conditions close to $T_c$.
Before an excitation gap ever develops, cooling through the superfluid transition \emph{at a finite rate} leads to decoupling of long-lived thermal fluctuations of Cooper pairs just above $T_c$ from the neutron bath temperature that is tied to the cooling rate, i.e. Cooper pair fluctuations fall out of equilibrium with the bath temperature $T(t) = T_c - t/\tau_Q$ within a temperature window $T\in\{T_c-\delta_T, T_c+\delta_T\}$.
The nonequilibrium window, $\delta_T$, depends on the quench rate, $1/\tau_Q$, which is the neutrino cooling rate at $T_c$, as well as the microscopic pair correlation time of order $\hbar/\kb T_c$.

For neutrino cooling governed by the Urca (modified Urca) process the quench rate, $1/\tau_{Q}$, is given by Eq.~\ref{eq-tau_Urca} (Eq.~\ref{eq-tau_mUrca}) evaluated at $T_c$.
Thus, for transition temperatures spanning $T_c\in\{0.5-5\,\times 10^9\,\mbox{K}\}$ the quench timescale for the Urca process spans, $\tau_Q\in\{4.0\times 10^{3},4.0\times 10^{-1}\}\,\mbox{sec}$ for $n=n_0$, $\mu_e=50\,\mbox{MeV}$, $M_n^*/M_n=1$ and $M_p^*/M_p=1$. The corresponding quench rates for modified Urca cooling are considerably slower, but still sufficiently fast to create the nonequilibrium conditions necessary for the generation of a large population of vortices and vortex loops upon cooling through $T_c$.

\onecolumngrid
\begin{center}
\begin{figure}[t!]
\begin{minipage}{0.8\textwidth}
\includegraphics[width=\textwidth]{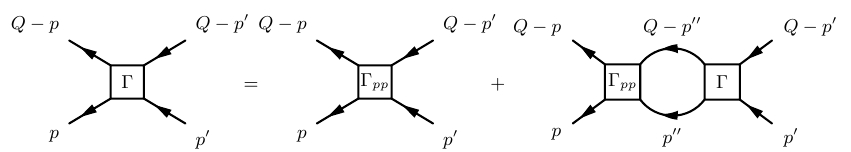}
\caption{Cooper Pair Propagator. $\Gamma_{pp}(p+Q,-p;p'+Q,-p')$ is the particle-particle irreducible vertex responsible for the binding of pairs of Fermions with nearly zero total four-momentum, $Q\ll(p_f,E_f)$.}
\label{fig-pair_propagator}
\end{minipage}
\end{figure}
\end{center}
\twocolumngrid

\section{Cooper Pair Fluctuations}
\vspace*{-5mm}

Inelastic scattering of neutron quasiparticles in degenerate normal nuclear matter have scattering rates of order 
\be
1/\tau_{nn}\approx (E_{f}/\hbar)\,N(0)^2\,\langle|\Gamma|^2\rangle_{\text{FS}}\,
\left(\varepsilon/E_{f_n}\right)^2
\,,
\ee
for an incident quasiparticle with excitation energy $\varepsilon$, where $N(0)=M_n^*k_{f_n}/2\pi^2\hbar^2$ is the density of states at the Fermi energy, $E_{f_n}$, and $\Gamma$ is the transition amplitude for binary collisions of neutron quasiparticles with momenta and excitation energies near the Fermi surface.
This timescale is of order $\tau_{nn}\approx 10^{-7}\,\mbox{ns}\ll\tau^{\urca}$ for thermal quasiparticles with $\varepsilon\sim\kb T \simeq 10^{-3}E_{f_n}$.
Thus, normal degenerate neutron matter remains essentially in local equilibrium during cooling even for Urca processes.
Also note that the limited phase space, $\sim(\varepsilon/E_{f_n})^2$, for binary collisions at low energy also means that neutron quasiparticles are long-lived excitations: $\hbar/\tau_{nn}(\varepsilon) \lll |\varepsilon|$.

In this context the Cooper instability occurs for an arbitrarily weak attractive interaction in a bandwidth $|\varepsilon| < \omega_c \ll E_{f_n}$ around the Fermi energy for pairs of quasiparticles with center of mass energy and momentum $Q=(\omega,\vQ)\rightarrow 0$. 
For homogeneous nuclear matter under equilibrium conditions at fixed density, $n$, and temperature, $T$, the transition to a spatially homogeneous superfluid of pairs of bound neutron quasiparticles is preceded by long-lived, long-wavelength 
Cooper pair fluctuations as $T\rightarrow T_c$.

The propagator for these ``precursor pairs'' is governed by a Bethe-Salpeter equation for the amplitude of neutron quasiparticle pairs with total energy and momentum in the center of mass channel, $Q=(i\omega_m,\vQ)$, shown diagrammatically in Fig.~\ref{fig-pair_propagator},

\onecolumngrid
\vspace*{-5mm}
\be
\Gamma(p,p';Q) = \Gamma_{pp}(p,p';Q) 
-\onehalf\,\int dp''\,
\Gamma_{pp}(p,p'';Q)\,G(p''+Q)\,G(-p'')\,\Gamma(p'',p';Q)
\,,
\ee
\twocolumngrid
%
where $p=(i\eps_n,\vp)$, $p'=(i\eps_n',\vp')$, $\eps_n=(2n+1)\pi T$ and $\int dp''(\ldots) \equiv T\sum_{\eps_{n''}}\int d^3p''/(2\pi)^3(\ldots)$.\footnote{Here I set $\kb=\hbar=1$ and restore normal dimensions when needed. The factors of $(-1)$ and $\tinyonehalf$ come from the Feynman rules for vertices and equivalent Fermion lines, respectively.}

The particle-particle irreducible vertex, $\Gamma_{pp}$, provides the attraction that binds neutron quasiparticles into Cooper pairs. Both direct nuclear forces and induced interactions mediated by the nuclear medium contribute to $\Gamma_{pp}$.~\cite{bab73,sch03}
Here I consider spin-singlet pairing fluctuations in the weak-coupling limit, $\kb T_c\ll \omega_c\ll E_{f}$.
The propagator for spin-triplet, p-wave Cooper pair fluctuations has the same structure, but includes a $3\times 3$ multiplicity of modes of pairing fluctuations.~\cite{lin21}

The pairing interaction, $\Gamma_{pp}$, is to good approximation given by a pseudopotential confined to the low-energy shell of attraction near the Fermi surface, $|\eps|\le \omega_c \ll E_{f_n}$ and $Q=0$,

\be
\Gamma_{pp}(p,p')
\approx
\gamma_{pp}(\hat{\vp}\cdot\hat{\vp}')\,
\Theta(\omega_c-|\eps_n|)\,\Theta(\omega_c-|\eps_n'|)
\,.
\ee
For rotationally invariant interactions we can expand in the basis of spherical harmonics,
\be
\gamma_{pp}(\hat\vp\cdot\hat\vp')
=
\sum_{l=0}^{\text{even}}\,\gamma_{l}\,
\sum_{m=-l}^{+l}\,\cY_{lm}(\hat\vp)\,\cY^*_{lm}(\hat\vp')
\,,
\ee
where $\gamma_l$ is the pairing interaction in angular momentum channel $l$.
The Cooper instability is then determined by the channel with the most attractive interaction. For the case of $^1S_0$ pairing, $\gamma_0$ is the most attractive channel and the orbital state of the Cooper pairs is s-wave. 
The resulting propagator for precursor $^1S_0$ pairs can also be expressed in the low-energy shell as $\Gamma(p,p';Q)\approx \gamma(Q)\,\Theta(\omega_c-|\eps_n|)\,\Theta(\omega_c-|\eps_n'|)$, with 
\be
\gamma(Q)=\frac{\gamma_0}{1+\tinyonehalf\gamma_0\,\Pi(Q)}
\,,
\ee
\be
\mbox{where} 
\,\,
\Pi(Q)=T\sum_{\eps_n}^{|\eps_n|\le\omega_c}\int\dthree{p}\,G(p+Q)\,G(-p)
\ee
is the pair susceptibility, i.e. the intermediate pairing state shown in Fig.~\ref{fig-pair_propagator}.
In the low-energy shell the neutron propagator is given by 
\be 
G(p) = \frac{1}{i\eps_n - (|\vp|^2/2M_n^* - \mu)}
\,,
\ee
with $(|\vp|^2/2M_n^* - \mu)\equiv\xi_{\vp}\approx v_f(|\vp|- p_f)$, where $\vv_{\vp}=v_f\hat{\vp}$ with $v_f= p_f/M_n^*$ is the Fermi velocity and $M_n^*$ is the effective mass of neutron quasiparticles in nuclear matter. Thus, we can evaluate $\Pi(Q)$ as 
\ber
\hspace*{-5mm}
\Pi(Q)
&=&
N(0)\,T\sum_{\eps_n}^{|\eps_n|\le\omega_c}\,\int\frac{d\Omega_{\vp}}{4\pi}\,\int\,d\xi_{\vp}
\nonumber\\
&\times& 
\frac{1}{\xi_{\vp}+\vv_{\vp}\cdot\vQ-i(\eps_n+\omega_m)}\times\frac{1}{\xi_{\vp}+i\eps_n}	
\,.
\label{eq-pair_bubble}
\eer

For $Q=({\bf 0},0)$ the pair propagator is regulated by the cutoff in the bandwidth of the attractive pairing interaction, $\Pi({\bf 0},0)=N(0)\,\ln(1.13\,\omega_c/T)$. 
The resulting static Cooper pair propagator for $\vQ={\bf 0}$,
\be
\gamma({\bf 0},0)
=
\frac{\gamma_0}{1+\onehalf\gamma_0\,N(0)\,\ln(1.13\,\omega_c/T)}
\,,
\ee
%
diverges for an attractive interaction, $\gamma_0 < 0$, at temperature, $T_c=1.13\,\omega_c\,e^{-1/\lambda_0}$, where $\lambda_0 = \tinyonehalf N(0)|\gamma_0|$ is the dimensionless pairing interaction, which is the well known Cooper instability.~\cite{AGD2}

The onset of superfluidity is preceded by long-lived fluctuations of Cooper pairs which are described by the long-wavelength, low-frequency Cooper pair propagator. For $|\vQ|<2\pi T_c/v_f$ and $|\omega_m|<T_c$ the leading corrections to the pair susceptibility from Eq.~\ref{eq-pair_bubble} are
%
\be
\hspace*{-2mm}
\Pi(Q)
\ns=\ns
N(0)
\Bigg\{
\ns
\ln\left(\frac{1.13\,\omega_c}{T}\right)
\ns-\ns
\frac{\pi}{8 T}\,\omega_m
\ns-\ns
\frac{7\zeta(3)v_f^2}{48\pi^2T^2}|\vQ|^2
\ns\Bigg\}
.
\ee

Thus, for temperatures $T\approx T_c$ the propagator for Cooper pair fluctuations becomes, 
\be
\hspace*{-2mm}
\gamma^{\text{R}}(\vQ,\omega) = -\frac{2}{N(0)}\,
\left\{\frac{1}{\ln(T/T_c) - i\omega\tau_0 + |\vQ|^2\xi_0^2}\right\}
\,,
\label{eq-pair_propagator}
\ee
upon analytical continuation to the real axis, $i\omega_m\rightarrow \omega + i0$, to obtain the retarded propagator for Cooper pair fluctuations, and is valid for $\omega\tau_0<1$ and $|\vQ|\xi_0<1$. The length and timescales defining the propagator are given by~\footnote{The Riemann zeta function $\zeta(3)\simeq 1.20205$~[\onlinecite{abramowitz84}]}
\ber
\xi_0 = \sqrt{\frac{7\zeta(3)}{12}}\,\frac{v_f}{2\pi T_c}
\,,\qquad
\tau_0 = \frac{\pi}{8 T_c}
\,.
\eer

For $T>T_c$ the pole is in the lower half of the complex $\omega$ plane. Thus, the retarded propagator in the time-domain is obtained from the pole 
\be
\gamma^{\text{R}}(\vQ,t)=\left(\frac{-2}{N(0)\tau_0}\right)\,
e^{-t/\tau(\vQ,T)}
\,;\quad t>0
\,,
\label{eq-pair_propagator-time_domain}
\ee
where the pair fluctuation lifetime is given by
\be
\tau(\vQ,T) = \frac{\tau_0}{(T/T_c - 1) + |\vQ|^2\xi_0^2}
\,.
\ee
Thus, $\vQ=0$ pairs become very long-lived as $T\rightarrow T_c^+$. The spatial scale for long-lived fluctuations is set by the correlation length, $\xi(T)=\xi_0/[T/T_c-1]^{\tinyonehalf}$, i.e. fluctuations with $|\vQ|^{-1}\gg\xi(T)$ are long-lived near $T_c$, while shorter wavelength modes decay more rapidly.

\vspace*{-5mm}
\section{Kibble-Zurek Freeze-Out}
\vspace*{-5mm}

For temperatures just below $T_c$ the Cooper pair propagator in Eq.~\eqref{eq-pair_propagator-time_domain} exhibits exponential growth of long-wavelength Cooper pairs, i.e. modes with $|\vQ|^{-1}\gg\xi(T)$, where 
\be
\xi(T)=\frac{\xi_0}{|1 - T/T_c|^{\tinyonehalf}}
\,,
\ee
is the superfluid correlation length. This leads to the formation of a pair condensate, i.e. a macroscopic population of Cooper pairs. The $\vQ=0$ mode grows at the slowest rate, with growth cutoff by the formation of a macroscopic condensate density after a time of order
\be
\tau(T) = \frac{\tau_0}{1-T/T_c}
\,.
\ee

For a thermodynamic transition the temperature drops infinitesimally slowly through $T_c$. Thus, the pairing fluctuations remain in equilibrium and decay as $T\rightarrow T_c^+$ for all finite $\vQ$ even though the dynamics of the longer wavelength fluctuations slows dramatically for temperatures close to the transition.
As a result the improbable fluctuation with $Q\rightarrow 0^+$, with a diverging lifetime, can grow adiabatically into a homogeneous condensate once $T$ passes $T_c$.

However, in the case of a temperature quench, pairing fluctuations of various wavelengths created just above $T_c$ do not all decay as the temperature passes through the transition. Below $T_c$ each of the surviving fluctuations begins to grow in magnitude.
Furthermore, these fluctuations form an ensemble of localized, spatially distributed and causally independent patches of precursor condensate.

Consider the space-time evolution of a local Cooper pair fluctuation of amplitude $\cA=|\cA|e^{i\vartheta}$ centered at $\vR=0$ at $t=0$ for $T=T_c$. The growth of this fluctuation for $t>0$ is
%
\ber
\hspace*{-3mm}
\gamma^{\text{R}}(\vR,t) 
\ns=\ns\cA\,
e^{+t/\tau(T)}
\int\dthree{Q}\,e^{i\vQ\cdot\vR}\,e^{-|\vQ|^2\xi_0^2\,t/\tau_0}
\nonumber\\
\ns=\ns\frac{\cA}{\left(2\xi_0\sqrt{\pi t/\tau_0}\right)^{3}}\,
\exp{\displaystyle{\left[\frac{t}{\tau(T)}-\frac{R^2}{4\,\xi_0^2\,t/\tau_0}\right]}}
\,.
\eer

Thus, the local fluctuation grows in amplitude and \emph{expands} 
with radial extent given by
\be
\hspace*{-3mm}
R_{\text{B}}(t)
=2\xi_0
\left(\frac{t}{\tau_0}\right)
\left(\frac{t}{\tau_Q}\right)^{1/2}
\ns\ns,
\ee
where the latter expression is the radius of the fluctuation ``bubble'' as a function of the time after the initial quench below $T_c$, which depends on both the microscopic timescale, $\tau_0$, as well as quench timescale, $\tau_Q$.

Now if we consider the fact that different regions of the liquid develop condensate patches of size $R_{\text{B}}(T)$ after $t$ seconds of cooling below $T_c$, these regions are causally disconnected, with the random phase of each patch belonging to a distinct element of the degeneracy manifold. The initial space-time evolution continues \emph{until} the ensemble of patches grows to sufficient size to establish global phase coherence. 
This is achieved when the patch size, $R_{\text{B}}(t)$, becomes of order the coherence length, $\xi(T(t))$, at which time the distinct phases of the individual patches begin to align via local currents that anneal out inhomogeneities in the condensate phase.
The time at which the distinct patches begin interacting to form a causally connected condensate structure is the Kibble-Zurek freeze-out time, $\hat{t}$, defined here by
\be
R_{\text{B}}(\hat{t})=2\xi(T(\hat{t}))
\quad\leadsto\quad
\hat{t}=\sqrt{\tau_Q\tau_0}
\,.
\ee

This is the Zurek's result for the freeze-out time based on mean-field exponents, $\nu=\nicefrac{1}{2}$ and $z=2$.\cite{zur85}

%
\begin{center}
\begin{figure}[t!]
\begin{minipage}{0.495\textwidth}
\includegraphics[width=\textwidth]{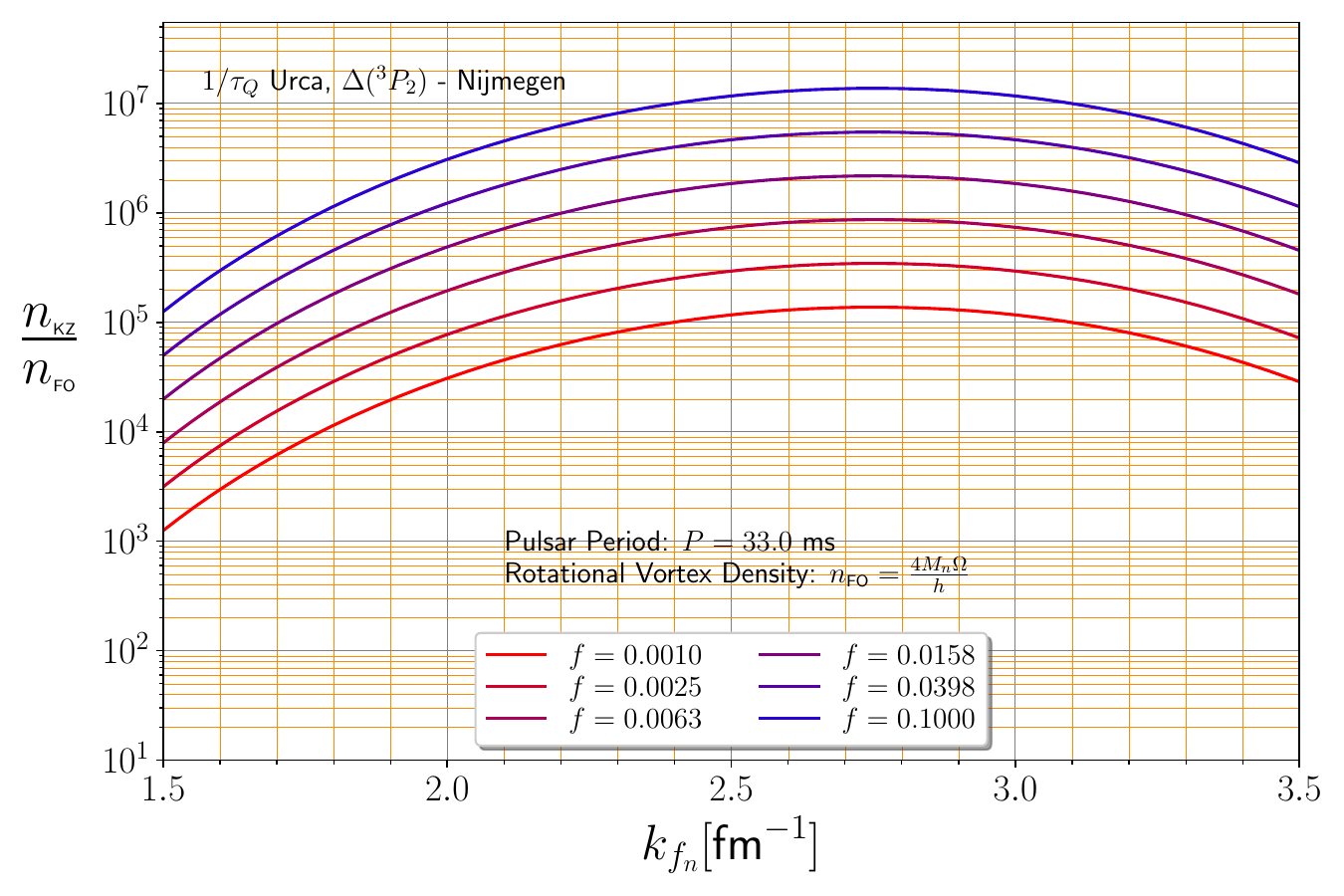}
\end{minipage}
\begin{minipage}{0.495\textwidth}
\includegraphics[width=\textwidth]{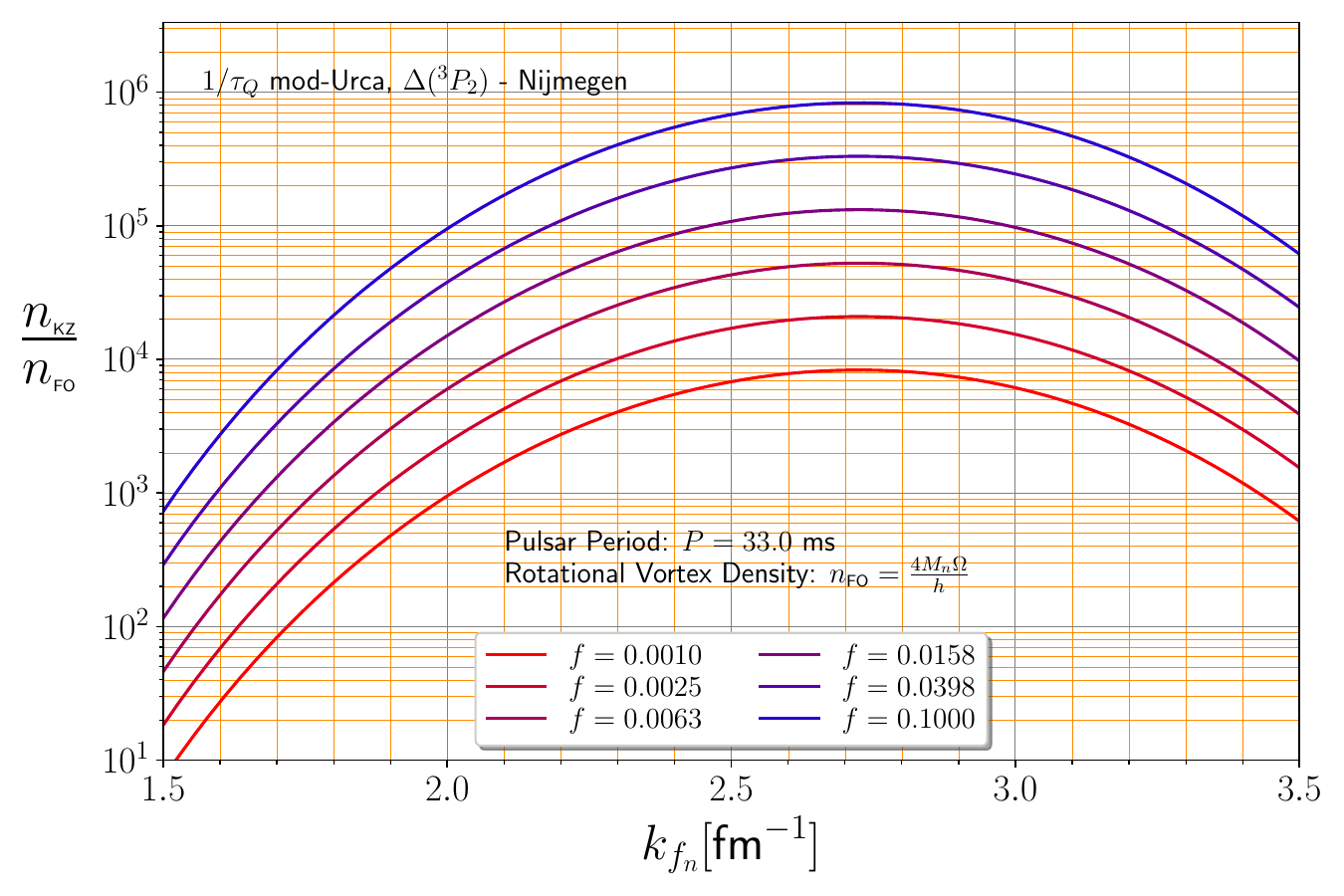}
\end{minipage}
\caption{The areal density of vortices generated by the KZM as a function of neutron Fermi wavenumber $k_{f_n}$, normalized by the areal density of equilibrium vortices for co-rotation at the Crab pulsar rotational velocity.
The fraction $f$ of $\pm 2\pi$ phase windings at freeze-out is varied between $10^{-3} - 10^{-1}$.
The quench rate through $T_c$ is obtained from direct Urca neutrino cooling (top panel) and modified Urca neutrino cooling (bottom panel). The \pwave\ gap and $T_c$ are based on the Nijmegen-I potential from Ref.~\onlinecite{bal98}.}
\label{fig-nKZ_vs_kfn-Urca+mod-Urca+Nijmegen}
\end{figure}
\end{center}

\vspace*{-5mm}
\vspace*{-5mm}
\subsection{Vortex Formation \`a la Kibble-Zurek}\label{sec-vortex_formation-KZM}
\vspace*{-5mm}

Phase differences between connected patches of condensate anneal out to form larger regions of nearly uniform phase, \emph{unless} the phase winding connecting multiple patches, $(1/2\pi)\oint_{C}\,d\vl\cdot\grad\vartheta\in\{\pm 1,\pm 2,\ldots\}$ enforces a \emph{phase vortex} embedded in the precursor connected condensate.
Regions of linear size of order 
\be
\hat{\xi}\equiv\xi(T(\hat{t}))
=
\xi_0\,\left(\frac{\tau_Q}{\tau_0}\right)^{\frac{1}{4}}
\,,
\ee
or larger will host phase variations that cannot be annealed out, i.e. they are \emph{non-removable} because the winding number
\be\label{eq-winding_number}
N_{\text{C}}=\frac{1}{2\pi}\oint_{\text{C}}d\vl\cdot\grad\vartheta\in\{\pm 1,\pm 2\,\ldots\}
\,,
\ee 
of the connected condensate for these patches provides a topological constraint protecting a node of the condensate amplitude enclosed within {\small C}.
Thus, for line defects in 3D, such as phase vortices, a fraction $f$ of regions of minimal area $\hat{\xi}^2$ in \emph{any} two-dimensional cross section of $\mathbb{R}^3$ will host phase vortices. The mean areal density of phase vortices at freeze-out is then
\be
n^{\text{KZ}}_v 
=f\,\hat\xi^{-2}
=\frac{f}{\xi_0^2}\,\left(\frac{\tau_0}{\tau_Q}\right)^{\frac{1}{2}}
\,.
\label{eq-defect_density-KZM}
\ee
This is the scaling prediction for the areal density of phase vortices generated via the KZM.~\cite{zur00} Note that $f$ is not universal, but specific to the physical system and the space-time dynamics governing the order parameter. Numerical simulations for KZ generation of topological defects in 1D, 2D and 3D nonequilibrium phase transitions report values of $f\in\{1.0, 0.0001\}$.~\cite{zur00}
Equation~\eqref{eq-defect_density-KZM} implies a mean distance between singular cores of phase vortices in any two-dimensional cut through $\mathbb{R}^3$,
\be
d^{\text{KZ}}_v=\hat\xi/\sqrt{f}
\,.
\label{eq-defect_spacing-KZM}
\ee

In Fig.~\ref{fig-nKZ_vs_kfn-Urca+mod-Urca+Nijmegen} we show the results for the areal density of vortices generated by the KZM based on neutrino cooling in the inner core via the Urca process. The upper (lower) panel is for the Urca (modified Urca) process and the \pwave\ gap based on the Nijmegen I potential shown in Fig.~\ref{fig-Delta_vs_kfn} from Ref.~\onlinecite{bal98}. 
The vortex density is normalized by the areal density of vortices that establishes co-rotation with the crust and normal matter at rotation speed $\Omega$ corresponding to that of the Crab pulsar.

The first observation is the enormous density of vortices that are generated by the KZM based on the Urca cooling rate for a \pwave\ gap varying from $0.1 - 1.4$ MeV in the core.
The second observation is that even the slower modified-Urca rate generates a huge nonequilibrium vortex density, and that remains the case even for smaller \pwave\ gaps.

\begin{center}
\begin{figure}[t!]
\begin{minipage}{0.495\textwidth}
\includegraphics[width=\textwidth]{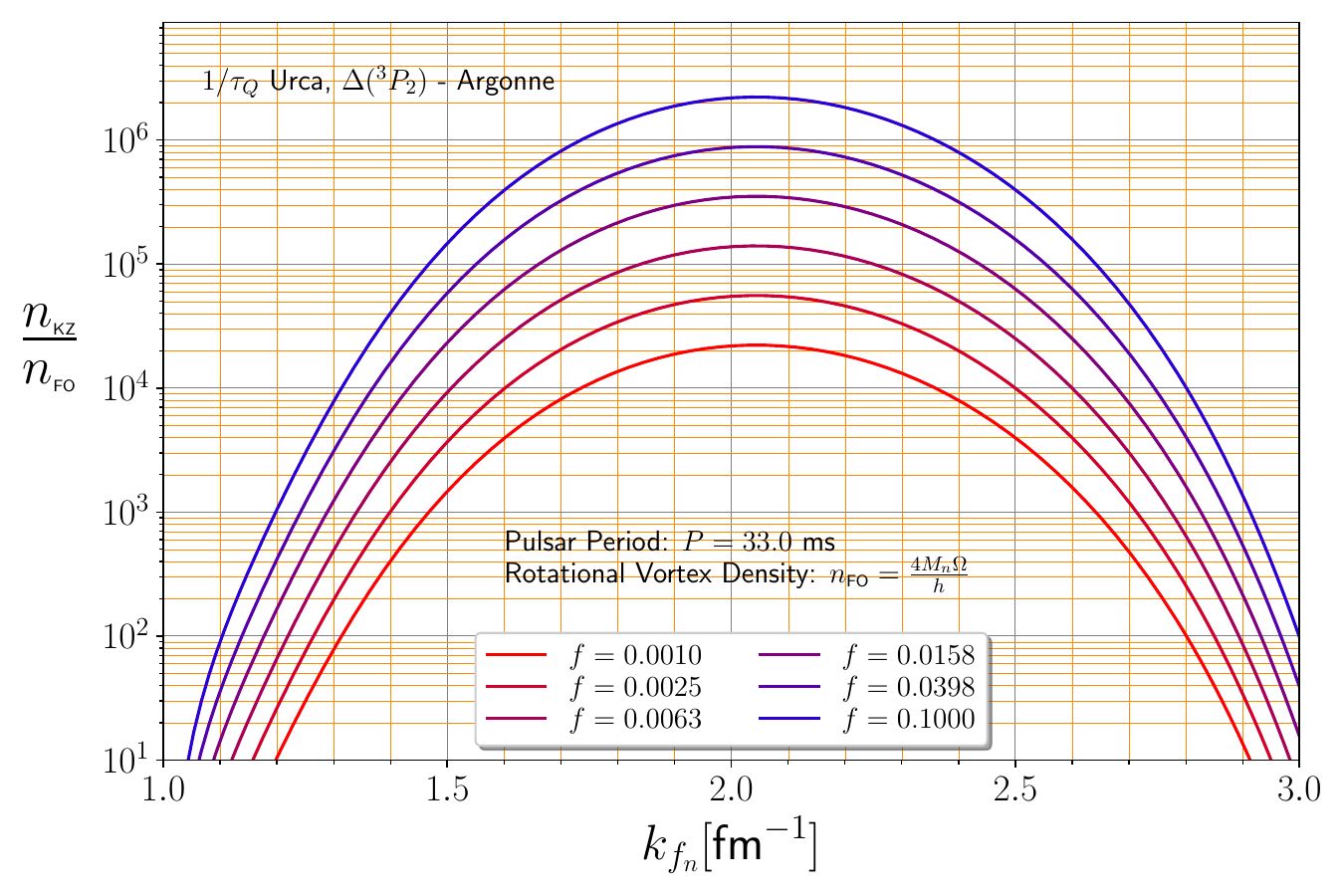}
\end{minipage}
\begin{minipage}{0.495\textwidth}
\includegraphics[width=\textwidth]{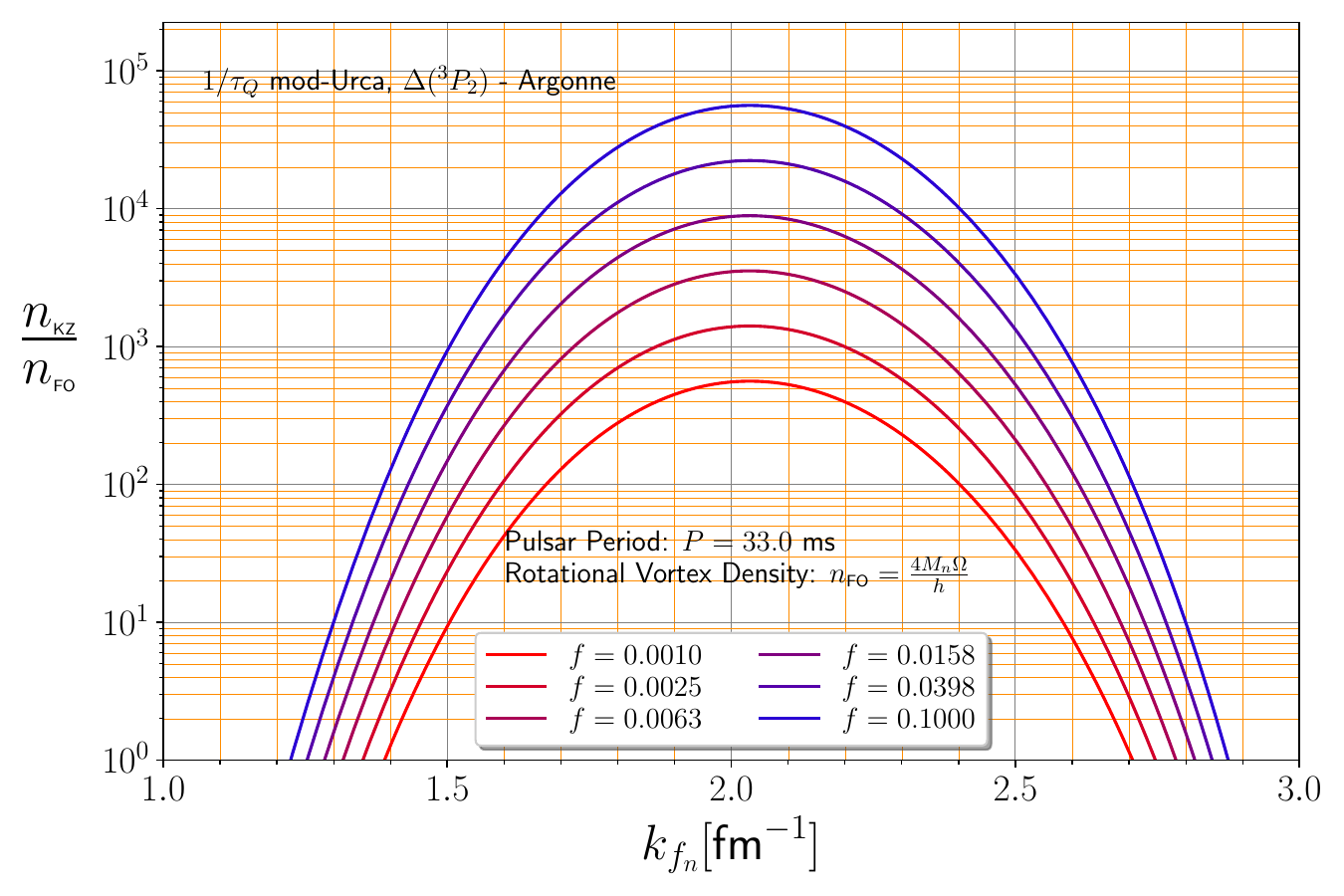}
\end{minipage}
\caption{
Same caption as that in Fig.~\ref{fig-nKZ_vs_kfn-Urca+mod-Urca+Nijmegen} 
except the \pwave\ gap and $T_c$ are based on the Argonne potential from Ref.~\onlinecite{bal98}.}
\label{fig-nKZ_vs_kfn-Urca+mod-Urca+Argonne}
\end{figure}
\end{center}

Theoretical models of the core with a significantly reduced \pwave\ gap also exhibit large vortex densities, reduced by roughly an order or magnitude and reduced region of the inner core as shown in Fig.~\ref{fig-nKZ_vs_kfn-Urca+mod-Urca+Argonne}. The central message is that the KZM based on the more conservative estimate of the pairing gap and modified Urca process predict large vortex densities over most of the inner core. 

Thirdly, for the outer core and inner crust the modified Urca process, even for the smaller \swave\ pairing gap shown in Fig.~\ref{fig-Delta_vs_kfn}, generates vortex densities comparable to those shown in the lower panel of Fig.~\ref{fig-nKZ_vs_kfn-Urca+mod-Urca+Argonne} for the modified Urca cooling rate.

A fourth, and key, point is that the vortex areal density generated by the KZM for a nonequilibrium superfluid transition is the mean areal density for \emph{any 2D plane}. 
Thus, the KZM generates a highly inhomogeneous superfluid phase with a random entangled network of vortex lines and loops in 3D giving rise to a complex velocity distribution that evolves and coarsens over time.~\cite{zur00}
Such a state satisfies the basic definition of quantum turbulence, but the random vortex network generated by a KZM quench may differ in fundamental ways from numerical quantum turbulence studies based on the vortex filament model,~\cite{barenghi2024} or numerical simulations based on the Gross-Pitaevskii equation with energy injection at large scales and dissipation at short distances.~\cite{tsu08a} 

%

\vspace*{-15mm}
\subsection{Nematic Defects of a \pwave Superfluid}
\vspace*{-5mm}

In the inner core it is important to point out that the KZM will generate additional topologically stable defects that are unique to the \pwave\ ground state, and distinct from phase vortices responsible for quantized mass circulation.
The KZM operates on the degeneracy space of the \pwave\ ground state. In this case the ground state is defined by the traceless rank two tensor order parameter,~\cite{sau78}
\be
A_{\mu\nu} = \Delta\,e^{i\vartheta}\,
\left[\hat\vu_\mu\hat\vu_\nu + r\,\hat\vv_\mu\hat\vv_\nu -(1+r)\hat\vw_\mu\hat\vw_\nu\right]
\,,
\label{eq-3P2_biaxial_nematic}
\ee
where $\{\hat\vu,\hat\vv,\hat\vw\}$ is an orthonormal triad of unit vectors defining the spin and orbital 
coordinates of the ground state \pwave\ Cooper pairs. Indices $\mu,\nu\in\{x,y,z\}$ define the direction cosines of the triad with respect to a fixed Cartesian coordinate system, and $-\tinyonehalf \le r \le -1$ specifies the spin-orbit structure of the \pwave\ Cooper pairs. For $r=-\tinyonehalf$ the Cooper pairs are defined by a uniaxial tensor describing the orientation of the spin- and orbital states,
\be
A_{\mu\nu} = \Delta\,e^{i\vartheta}\,\left(\hat\vu_\mu\hat\vu_\nu - \tinyonethird\,\delta_{\mu\nu}\right)
\,.
\ee
This is the order parameter of a {\sl superfluid nematic liquid crystal}, and it is the ground state within weak-coupling BCS theory for \pwave\ pairing for core magnetic fields typical of radio pulsars, 
$B\lesssim B_c\simeq 10^{12}\,\mbox{G}$.~\cite{sau78,sau80}
In particular, the nematic phase defines \pwave\ Cooper pairs with $\hat\vu\cdot\vJ=0$.

What is relevant in terms of the generation of topological defects by the KZM is the degeneracy space of the nematic phase. For closed circuits $\cC$ in ${\mathbb R}^3$, embedded phase vortices are identified by mappings in which $\vartheta \xrightarrow[]{\cC}\vartheta + 2\pi\,N$ with winding numbers $N\in\{\pm 1,\pm 2,\ldots\}$, although only $N=\pm 1$ vortices are reported in simulations of quench-generated vorticity.~\cite{ant99,glu25}

The \pwave\ nematic supports additional line defects corresponding to rotations of the nematic axis by $\pi$ around closed circuits, $\hat\vu\xrightarrow[]{\cC}-\hat\vu$, that leave the nematic order parameter invariant. These defects are the \pwave\ Cooper pair analog of the well known {\sl disclination lines} of nematic liquid crystals.~\cite{deGennes93}

Based on a time-dependent extension of the Ginzburg-Landau theory for \pwave\ superfluids~\cite{sau80} along the same lines as that developed for superfluid \He~\cite{sau17,hin24,glu25} I expect disclination lines and loops to be generated by the KZM in roughly equal measure to those of phase vortices.
However, the postfreeze-out dynamics and coarsening of these distinct populations of topological defects may differ in the long time limit. 
The reason is that once long-range phase coherence is established, co-rotation of the superfluid component with the crust and normal components is enforced through the Feynman-Onsager constraint on the population of vortices and anti-vortices.
By contrast there is no such constraint on the population of the disclination lines and loops. Furthermore, it is an open question as to the role of interactions between the two populations of line defects, i.e. whether or not vortices pin to disclinations, and if so what the implications are for the the decay of quantum turbulence.

In addition to line defects the \pwave\ nematic phase supports point defects (``hedgehogs'') of the nematic axis, which I also expect to be generated by the KZM.~\footnote{The classification of topologically stable defects based on homotopy groups in \pwave\ superfluids is discussed in Refs.~\onlinecite{sau80,muz80}.}
The role of point defects in the postfreeze-out coarsening of the inhomogeneous \pwave\ phase, and the decay of quantum turbulence are interesting problems for future study.

Finally, I note that in magnetars the interior magnetic field of the neutron star 
is sufficiently strong that the nuclear Zeeman energy of the \pwave\ neutron pairs competes with 
the condensation energy. The ground state that minimizes the Zeeman energy is the biaxial nematic 
phase with $r=-1$,
\be
A_{\mu\nu} = \Delta\,e^{i\vartheta}\,
\left(\hat\vu_\mu\hat\vu_\nu - \hat\vv_\mu\hat\vv_\nu\right)
\,,
\label{eq-biaxial_nematic_OP}
\ee
and $\hat\vu\cdot\vB=0$, $\hat\vv\cdot\vB=0$.~\cite{sau80,muz80,mas22}
Thus, in addition to singly quantized phase vortices the high-field biaxial nematic superfluid 
also hosts topologically stable {\sl mixed vortex-disgyration} line defects in which 
$\vartheta \xrightarrow[]{\cC}\vartheta + \pi$, 
$\hat\vu\xrightarrow[]{\cC}\hat\vv$ and
$\hat\vv\xrightarrow[]{\cC}-\hat\vu$, 
which leaves $A_{\mu\nu}$ in Eq.~\eqref{eq-biaxial_nematic_OP} invariant.
These line defects have quantized mass circulation of half that of singly quantized phase vortices, 
and line energies approximately half that of singly quantized vortices in weak-coupling BCS 
theory.~\cite{sau80,muz80} 
The internal structure and energetics of these ``half-quantum'' vortices is discussed in 
detail in Ref.~\onlinecite{mas22}.

Thus, if neutrino cooling is as efficient in magnetars as in radio pulsars and the interior of magnetars supports \pwave\ superfluidity then quantum turbulence generated by neutrino cooling through the transition is expected to involve both singly quantized vortices and half-quantum vortices with potentially important implications for the evolutionary dynamics of the superfluid core.

\vspace*{-5mm}
\subsection{Inhomogeneous Quench}\label{sec-inhomogeneous_KZM}
\vspace*{-5mm}

The results for vortex generation by the KZM presented in Sec.~\ref{sec-vortex_formation-KZM} are based on a locally homogeneous quench, and represent the maximum defect density at any time post KZ freeze-out. 
The transition temperature and neutrino cooling rate vary throughout the interior on length scales of several kilometers.
Thus, the nonequilibrium transition first onsets in regions of maximum $T_c$ for the \pwave\ (\swave) superfluid in the inner (outer) core, which is also the region of maximum rate of neutrino emission, then spreads radially to neighboring regions with decreasing $T_c$.
If the speed of the transition front is fast compared to the fluctuation growth speed,
\be
v_{\text{B}}=v_0\,\sqrt{1-T(t)/T_c} = v_0\,\sqrt{t/\tau_Q}
\,,
\label{eq-bubble_speed}
\ee
where $v_0=3\xi_0/\tau_0$, the homogeneous KZM proceeds without restriction 
by a slow moving transition front, and provides an accurate estimate of the topological defect 
density as a function of the local freeze-out time.~\cite{del13}

To estimate the speed of the transition front note that
the gap profiles versus $k_{f_n}$ in Fig.~\ref{fig-Delta_vs_kfn} can be converted to profiles 
of $T_c$(\swave) and $T_c$(\pwave) as a function of radius, $T_c(r)$, with local maxima in the 
inner (\swave) and outer (\pwave) regions of the core of the neutron star. The radius of curvature 
at the maximum is $R_c\sim 1-3\,\mbox{km}$.  Thus, for the purpose of an analytical estimate of the 
transition front I assume a parabolic model of the superfluid dome centered at $r=R_{\text{max}}$, is
$T_c(x) = T_c\,(1-x^2/R_c^2)$, 
where $x\equiv r-R_{\text{max}}$ and $T_c$ is the maximum transition temperature, 
and a uniform quench rate $\tau_Q$.

The quench through the superfluid transition dome is governed by
$\varepsilon(x,t) = 1 - T(t)/T_c(x)$. 
The transition front, $\varepsilon(x,t)=0$, spreads as
\be
x(t)=R_c\,\left(\frac{t}{\tau_Q}\right)^{\tinyonehalf}\,,\quad t\ge 0
\,,
\ee
with speed,
\be
u_F(t) = \frac{R_c}{2\tau_Q}\,\left(\frac{\tau_Q}{t}\right)^{\tinyonehalf}
\,.
\ee
The speed of the transition front is initially divergent (since $\partial T_c/\partial x\vert_{x=0} = 0$), 
then gradually slows as $T_c(x)$ decreases. For the pairing fluctuations that begin to grow as the quench 
drops below the maximum at $T_c$ the fluctuation
growth speed is given by Eq.~\eqref{eq-bubble_speed}. Thus, 
\be
\frac{u_{\text{F}}(t)}{v_\text{B}(t)}=\nicefrac{1}{2}\,\frac{R_c}{v_0\,t}
\,.
\label{eq-velocity_ratio}
\ee
The ratio in Eq.~\eqref{eq-velocity_ratio} exceeds unity for 
$t<t_{\star}\equiv\tinyonehalf R_c/v_0 > 1$, with a corresponding radial spread of
$x_{\star}=x(t_{\star})=R_c\,\left(t_{\star}/\tau_Q\right)^{\tinyonehalf}$.
For the \pwave\ dome centered at $k_{f_n}=2.8\,\mbox{fm}^{-1}$ with maximum $\Delta=1.4\,\mbox{MeV}$ and 
estimated radius of curvature of $R_c = 2.5\,\mbox{km}$, $M^*_p/M_p = 0.7$, $\mu_e=100.0\,\mbox{MeV}$ and 
Urca cooling, gives $t_{\star}\simeq 5.8\times 10^{-6}\,\mbox{s}$ and $x(t_{\star})\simeq 62.3\,\mbox{m}$. 
Thus, $5\%$ of the superfluid dome goes through the superfluid transition in $5.8\,\mu\mbox{s}$ while
the transition front velocity exceeds the \emph{initial} fluctuation growth velocity.

As the front continues to expand newly generated pairing fluctuations just inside the front at 
$x_{<}=x(t-\Delta t)\simeq x(t) - u_F(t)\,\Delta t$ with $\Delta t\le \hat{t} \ll t$, have much 
slower growth speeds, 
\be
v_{\text{B}}(t|x_<) = v_0 \sqrt{1 - \frac{T_c(1-t/\tau_Q)}{T_c(1-x_<^2/R_c^2)}}
\simeq 
v_0\,\left(\frac{\Delta t}{\tau_Q}\right)^{\tinyonehalf}
\,,
\ee
than the instantaneous speed of the front. In particular, 
\be
\hspace*{-3mm}
\frac{u_{\text{F}}(t)}{v_\text{B}(t|x_<)}
=
\nicefrac{1}{2}\,\frac{R_c}{v_0}\,\frac{1}{\sqrt{t\,\Delta t}}
\gg 1\quad \mbox{for}\,\, t \ll \tau_Q\,,\Delta t\le\hat{t}
\,.
\ee
As a result generation of topological defects on the timescale of the local freeze-out time 
continue up to timescales of order $\tau_Q$, i.e. over essentially the entire transition dome
except close to the dome edge at $x(t\rightarrow\tau_Q)=R_c$ where 
the speed of the transition front slows dramatically.

It is worth emphasizing that, distinct from a homogenous quench, the population of topological 
defects undergoes \emph{inhomogeneous} coarsening even as topological defects are being generated 
as the front expands.
A proper treatment of the dynamics and generation of quantum turbulence, and the subsequent decay and 
coarsening of the vortex population requires 3D space-time simulations of the inhomogeneous 
quench dynamics, including spatial variations of the neutrino cooling rate as well as 
radial heat transport generated by the inhomgeneous neutrino cooling rate.
This in my view is an interesting future challenge.

\vspace*{-5mm}
\section{Conclusion and Outlook}\label{sec-summary}
\vspace*{-5mm}

Neutron stars born from a supernova event cool rapidly by beta decay reactions that generate neutrinos to transport energy out of the interior of the neutron star. Rapid cooling through the superfluid phase transition is shown to generate a highly inhomogeneous superfluid supporting quantum turbulence in the form of a high density of quantized vortices and vortex loops.
The results of this study are relevant for understanding the rotational dynamics of pulsars with superfluid cores, both at early times just after a nonequilibrium the phase transition, as well as longer timescales associated with the coarsening and decay of the quantum turbulent state. 
I considered the generation of quantized vortices by the KZM in the \swave\ condensate predicted to dominate the inner crust and outer core, as well as the \pwave\ condensate in the inner core. 
In the inner core the generation of topological defects and quantum turbulence also involves disclination lines and point defects intrinsic to the \pwave\ nematic superfluid phase.
Large scale numerical simulations of the postfreeze-out dynamics of the inhomogeneous phase of neutron star superfluids should provide new insight into the thermal evolution of the core as well as spin-down rates of pulsars, and perhaps novel sources of vortex pinning, annihilation and reconnection processes that may contribute to pulsar timing noise.

\noindent\underline{\sl Acknowledgements:} 
I thank Ali Alpar for many discussions on vortex dynamics in pulsars, including this work on Kibble-Zurek physics in superfluid neutron star matter.
Thanks to Mark Hindmarsh and Nobel Gluscevich for discussions and results on numerical simulations of nonequilibrium phase transitions in field theory models of early universe cosmology and BCS superfluids.
This research originated with the 1999 Les Houches workshop on {\sl Topological Defects and the Non-equilibrium Dynamics of Symmetry Breaking Phase Transitions}, and was presented in working groups on neutron stars at the Aspen Center for Physics.
My current research is supported by the Hearne Institute of Theoretical Physics and the Center for Computation and Technology at Louisiana State University.

\appendix

\vspace*{-5mm}
\section{Neutrino Cooling via Urca processes}\label{app-neutrino_cooling}
\vspace*{-5mm}

The neutrino luminosity generated by the direct Urca process based on a Fermi golden rule calculation for $n\rightarrow p+e^-+\bar\nu$ and $p+e^-\rightarrow n+\nu$ is given by~\cite{pet92}
\ber
L^{\urca}_{\nu} 
&=&
\frac{457\pi}{10,080}\,G_F^2\,\cos^2\theta_c\,\left(1+3\,g_A^2\right)\,
\frac{M_p^*\,M_n^*\,\mu_e}{\hbar^{10} c^5}
\nonumber\\
&\times&
\Theta\left(p_{f_e}+p_{f_p}-p_{f_n}\right)\,(\kb T)^6\,
\,,
\eer
where the Heaviside function is the kinematic constraint for the beta decay reaction to proceed with $p_{f_{e,p,n}}$ the Fermi momenta of degenerate electrons, protons and neutrons, respectively. 
The weak interaction coupling constant $G_F\simeq 8.96\times 10^{-5} MeV\,fm^3$, $\theta_c\simeq 0.239$ is the Cabbibo angle and $g_A \simeq -1.261$ is the axial-vector coupling.~\cite{lat91}
Fermi liquid corrections for the effective mass of neutron and proton excitations are included.~\cite{bal07}

The cooling rate for matter in the NS interior is governed by $C_v\,dT/dt = -L_{\nu}^{\urca}$, where the specific heat of degenerate nuclear matter above the superfluid transition is dominated by the neutron excitations,
\be
C_v = \frac{k_{f_n}\,M_n^*\,\kb}{3\hbar^2}\,\kb T
\,.
\ee
The cooling rate then reduces to Eq.~\eqref{eq-tau_Urca} with 
\ber
K^{\urca} 
&=& 
\frac{457\pi}{3360}\,G_F^2\,\cos^2\theta_c\,\left(1+3\,g_A^2\right)\,
\nonumber\\
&\times&
\frac{M_p c^2\,M_e c^2\,E_9^4}{(\hbar c)^8}\,
\nonumber\\
&\times&
\frac{M_p^*}{M_p}\,
\frac{\mu_e}{M_e c^2}\,
\frac{c}{k_0}\,
\label{eq-K_Urca}
\eer
where $k_0=(3\pi^2 n_0)^{\tinyonethird}$ $\simeq 1.644\,\mbox{fm}^{-1}$ and $E_9$ $\equiv \kb\,10^9\,\mbox{K}$ $\simeq 0.0862\,\mbox{MeV}$ and $\hbar c \simeq 197.33\,\mbox{MeV}$.

\vspace*{-3mm}
\section{Vortex Density due to Rotation}\label{sec-vortex_density-rotation}
\vspace*{-7mm}

The predictions of the vortex areal density and inter-vortex spacing generated by the KZM based on neutrino cooling through the superfluid transition are compared with the equilibrium vortex areal density for a uniformly rotating superfluid.
The neutron superfluid velocity is given by 
\be
\vv_s=\frac{\hbar}{2M_n}\grad\vartheta
\,, 
\label{eq-superfluid_velocity}
\ee
where $\vartheta$ is the local phase of the condensate wave function, and $2M_n$ is mass of a pair of neutrons. In the nonrelativisitic limit $\vv_s$ transforms as a velocity field under Galillean boosts. Importantly, $\vv_s$ is \emph{irrotational}, $\curl\vv_s=0$, except at singularities of the phase.
Such singularities are line defects, \emph{phase vortices}, with integer winding numbers given by Eq.~\eqref{eq-winding_number}. As a result Eq.~\eqref{eq-superfluid_velocity} implies quantization of superfluid circulation for any closed path in the condensate,~\cite{ons49,fey55}
$\oint_{\text{C}}\,d\vl\cdot\vv_s = N_{\text{C}}\,\frac{h}{2M_n}$,
where $K_n = h/2M_n$ is the quantum of circulation for pairs of neutrons.

For neutron matter rotating uniformly at angular speed $\Omega$ the superfluid component cannot co-rotate locally with the crust and normal fluid at $\vv_n=\vOmega\times\vr$.
However, the superfluid approximates co-rotation on a course-grained scale by hosting a uniform array of quantized vortices aligned with the axis of rotation, $\hat\vOmega$. 
In particular, for a superfluid interior of radius $R$, the total superfluid circulation, $N_v\,\frac{h}{2M_n}$, is spread over an area of $\pi R^2$.
The areal density of quantized line vortices, $n_v=N_v/\pi R^2$, that minimizes the kinetic energy in the rotating frame of the crust and normal fluid, and thus matches the circulation of the normal component, is given by the Feynman-Onsager constraint~\cite{ons49,fey55} 
\be
n_v^{\text{FO}}=\frac{4M_n\Omega}{h}
\,.
\ee
Thus, for the Crab pulsar with period $P=2\pi/\Omega\simeq33\,\mbox{ms}$ the areal density is $n_v^{\text{FO}}\simeq 0.002\,\mu\mbox{m}^{-2}$,
or a mean distance between the line vortices of
\be
d_v^{\text{FO}}=\nicefrac{1}{2}\sqrt{\frac{\hbar c\,Pc}{M_n c^2}}\approx 22\,\mu m
\,.
\ee
This is a macroscopic distance compared to the nuclear matter scale, $k_{f_n}^{-1}\sim \mbox{fm}$. Thus, co-rotation of the superfluid at typical pulsar rotation speeds requires only a very dilute array of rectilinear vortices spread out over distances of several kilometers.

%
%
\end{document}